\documentclass[twocolumn,twocolappendix]{aastex631}
\usepackage{bm}
\usepackage{amsmath,amsthm,amsfonts,amssymb,amscd}
\usepackage{threeparttable}
\usepackage[]{color}

\shorttitle{galaxy-halo misalignment}
\shortauthors{Xu, Jing \& Gao}

\begin{document}

\title{Mass Dependence of Galaxy-Halo Alignment in LOWZ and CMASS}

\correspondingauthor{Y.P. Jing}
\email{ypjing@sjtu.edu.cn}

\author[0000-0002-7697-3306]{Kun Xu}
\affil{Department of Astronomy, School of Physics and Astronomy, Shanghai Jiao Tong University, Shanghai, 200240, People’s Republic of China}
\affil{Institute for Computational Cosmology, Durham University, South Road, Durham DH1 3LE, UK}

\author[0000-0002-4534-3125]{Y.P. Jing}
\affil{Department of Astronomy, School of Physics and Astronomy, Shanghai Jiao Tong University, Shanghai, 200240, People’s Republic of China}
\affil{Tsung-Dao Lee Institute, and Shanghai Key Laboratory for Particle Physics and Cosmology, Shanghai Jiao Tong University, Shanghai, 200240, People’s Republic of China}

\author{Hongyu Gao}
\affil{Department of Astronomy, School of Physics and Astronomy, Shanghai Jiao Tong University, Shanghai, 200240, People’s Republic of China}

\begin{abstract}
We measure the galaxy-ellipticity (GI) correlations for the Slogan Digital Sky Survey DR12 LOWZ and CMASS samples with the shape measurements from the DESI Legacy Imaging Surveys. We model the GI correlations in an N-body simulation with our recent accurate stellar-halo mass relation from the Photometric object Around Cosmic webs (PAC) method. The large data set and our accurate modeling turns out an accurate measurement of the alignment angle between central galaxies and their host halos. We find that the alignment of central {\textit {elliptical}} galaxies with their host halos increases monotonically with galaxy stellar mass or host halo mass, which can be well described by a power law for the massive galaxies. We also find that central elliptical galaxies are more aligned with their host halos in LOWZ than in CMASS, which might indicate an evolution of galaxy-halo alignment, though future studies are needed to verify this is not induced by the sample selections. In contrast, central {\textit {disk}} galaxies are aligned with their host halos about 10 times more weakly in the GI correlation. These results have important implications for intrinsic alignment (IA) correction in weak lensing studies, IA cosmology, and theory of massive galaxy formation.
\end{abstract}

\section{Introduction} \label{sec:1}
Intrinsic alignment (IA) of galaxy shapes has been discovered for a long time \citep{2000ApJ...543L.107P,2002MNRAS.333..501B,2004MNRAS.353..529H,2006MNRAS.367..611M,2006MNRAS.369.1293Y,2009ApJ...694..214O,2009ApJ...694L..83O,2013ApJ...770L..12L,2022MNRAS.514.1077R}, and it has long been known as a main contamination to weak lensing measurement \citep{2000ApJ...545..561C,2004PhRvD..70f3526H}. Recently, IA has also been regarded as a promising cosmological probe \citep{2013JCAP...12..029C,2015JCAP...10..032S,2016PhRvD..94l3507C,2018JCAP...08..014K,2020MNRAS.493L.124O,2022PhRvD.106d3523O,2023ApJ...945L..30O,2023arXiv230202925K,2023NatAs.tmp..168X}. Moreover, IA of galaxies is highly related to their formation histories, since the strength of IA is found to depend on galaxy properties \citep{2015SSRv..193..139K,2015MNRAS.450.2195S,2021MNRAS.500.1895Z,2022MNRAS.514.1021J,2022arXiv221211319S}. Therefore, a deep understanding of IA of galaxies can benefit many fields in cosmology and astrophysics.   

IA of dark matter (DM) halos is well described by the linear alignment model \citep{2001MNRAS.320L...7C,2004PhRvD..70f3526H,2017ApJ...848...22X,2020MNRAS.493L.124O}, which are supported by N-body simulations \citep{2000MNRAS.319..649H,2000ApJ...545..561C,2002MNRAS.335L..89J,2020MNRAS.494..694O}. However, IA of galaxies is much less well understood,  due to its complicated dependences on morphology and color of galaxies  and on if they are centrals or satellites. Red and elliptical galaxies usually show higher IA strength and the signal is much lower for blue and disk galaxies \citep{2006MNRAS.369.1293Y,2016MNRAS.462.2668T,2020ApJ...904..135Y,2022MNRAS.514.1021J}. The luminosity and redshift dependence of IA of galaxies are also investigated in the literature \citep{2015MNRAS.450.2195S,2016MNRAS.461.2702C,2020MNRAS.491.4116B,2022PhRvD.106l3510H,2022arXiv221211319S}.

Most of the above studies directly measured the IA amplitude of galaxies and investigate its dependence on galaxy properties or redshift. Because the IA of halos are well understood, and galaxies are formed in halos, it would be very valuable to investigate how galaxies are aligned with their halos. Using the luminous red galaxies (LRG) from the Sloan Digital Sky Survey \citep[SDSS;][]{2000AJ....120.1579Y} DR6 and a large N-body simulation, \citet{2009ApJ...694..214O} and \citet{2009ApJ...694L..83O} first found that the misalignment angle between LRGs and their host DM halos can be described by a Gaussian distribution of  a dispersion $35^{\circ}$. Recently, \citet{2022PhRvD.106l3510H} investigated the misalignment angle distributions for Baryon Oscillation Spectroscopic Survey \citep[BOSS;][]{2015ApJS..219...12A} LOWZ and Dark Energy Survey \citep[DES;][]{2005astro.ph.10346T,2022PhRvD.105b3515S} samples, and they found nearly no luminosity or redshift dependence for central galaxies. In theory, \citet{2020MNRAS.491.4116B} found in hydrodynamic simulations that the IA of galaxies is governed by the IA of DM halos and the galaxy-halo misalignment angles. With galaxy-halo connection and galaxy-halo alignment, many works tried to model IA using the halo model \citep{2010MNRAS.402.2127S,2011JCAP...05..010B,2013MNRAS.436..819J,2021MNRAS.501.2983F,2022PhRvD.106l3510H}. An accurate observational determination of the misalignment angle for galaxies of different properties at different redshifts can serve as an indispensable ingredient for the IA halo model, and an important test for hydrodynamical simulations of galaxy formation.

In this work, using the high quality images of the DESI Legacy Imaging Surveys \citep{2019AJ....157..168D}, which covers the whole SDSS-III BOSS footprint, and with the recent accurate stellar-halo mass relation (SHMR) \citep{2023ApJ...944..200X} measured from Photometric object Around Cosmic webs (PAC) method \citep{2022ApJ...925...31X}, we measure the stellar mass, host DM halo mass and redshift dependence of the galaxy-halo misalignment angle for central elliptical and disk galaxies. The superior data set and the accurate modeling enable us to quantify the dependences of IA on the mass of galaxies and on the mass of their halos, and to find a significant evolution between the LOWZ and CMASS samples from redshift $z\approx 0.6$ to $z\approx 0.3$, though the difference between the samples should be considered. We will also show that central disk galaxies have a much weaker alignment with their host halos. We adopt the cosmology with $\Omega_m = 0.268$, $\Omega_{\Lambda} = 0.732$ and $H_0 = 71{\rm \ km/s/Mpc}$ throughout the paper.

\section{data and measurements}\label{sec:2}
To take the advantage of the accurate SHMR measurements from PAC \citep{2023ApJ...944..200X}, we use exactly the same samples as in \citet{2023ApJ...944..200X}. We use the SDSS-III BOSS DR12 LOWZ and CMASS spectroscopic samples\footnote{https://data.sdss.org/sas/dr12/boss/lss/} \citep{2015ApJS..219...12A,2016MNRAS.455.1553R} for two redshift ranges $0.2<z_s$\footnote{Throughout the paper, we use $z_s$ for spectroscopic redshift, $z$ for the $z$-band magnitude.}$<0.4$ and $0.5<z_s<0.7$ respectively. The galaxies are matched to the DR9\footnote{ https://www.legacysurvey.org/dr9/catalogs/} of the DESI Legacy Imaging Surveys to get the $grz$ band fluxes and shape measurements. Stellar masses of galaxies are then calculated using the spectral energy distribution (SED) code {\texttt{CIGALE}} \citep{2019A&A...622A.103B} with the \citet{2003MNRAS.344.1000B} stellar population synthesis models, the \citet{2003PASP..115..763C} initial mass function and the \citet{2000ApJ...533..682C} extinction law.

Then, we do central-satellite separation and morphology classification for the LOWZ and CMASS samples. As mentioned in \citet[See Figure 5]{2022ApJ...939..104X}, the photometric sample from the DESI Legacy Imaging Surveys with the photometric redshifts (photoz) calculated by \citet{2021MNRAS.501.3309Z} is suitable for studying the properties of massive galaxies ($>10^{11.0}M_{\odot}$). Thus, we use this photometric sample to select central galaxies from the LOWZ and CMASS samples. We calculate the stellar mass for the photometric sample in the same way but with photozs ($z_p$). Since the photozs have precision of $\sigma_{{\rm{NMAD}}}=0.02$ \citep{2021MNRAS.501.3309Z}, we regard the CMASS and LOWZ galaxies as centrals if there is no more massive photometric galaxies within $r_{{\rm{p}}}<1\ h^{-1}{\rm{Mpc}}$ and $|z_s-z_p|<0.1$. To ensure that all satellites are excluded from our samples, we have adopted conservative selection criteria, which may have also excluded some central galaxies, especially those at the lower end of the stellar mass range. Morphologies of galaxies are classified according to the S\'ersic index $n$ \citep{1963BAAA....6...41S}, with ellipticals having $n>2$.

The shapes of galaxies can be described by a two-component ellipticity, which is defined as
\begin{equation}
    e_{(+,\times)}=\frac{1-q^2}{1+q^2}(\cos{2\theta},\sin{2\theta})\,\,,\label{eq:1}
\end{equation}
where $q$ is the minor-to-major axial ratio of the projected shape, and $\theta$ is the angle between the major axis projected on to the celestial sphere and the projected separation vector pointing to a specific object. We use the {\texttt{shape\_e1}}\footnote{https://www.legacysurvey.org/dr9/catalogs/\#ellipticities} and {\texttt{shape\_e2}} in the catalog of DESI Legacy Imaging Surveys, which are measured using {\texttt{Tractor}} \citep{2016ascl.soft04008L}, as the shape measurements for each galaxies, and convert them to the ellipticity defined in Equation \ref{eq:1}. Following \citet{2009ApJ...694L..83O}, we assume that all the galaxies have $q=0$, which is equivalent to assuming that a galaxy is a line along its major axis. So, we only care about the orientations of the galaxies. 

\begin{figure*}
    \plotone{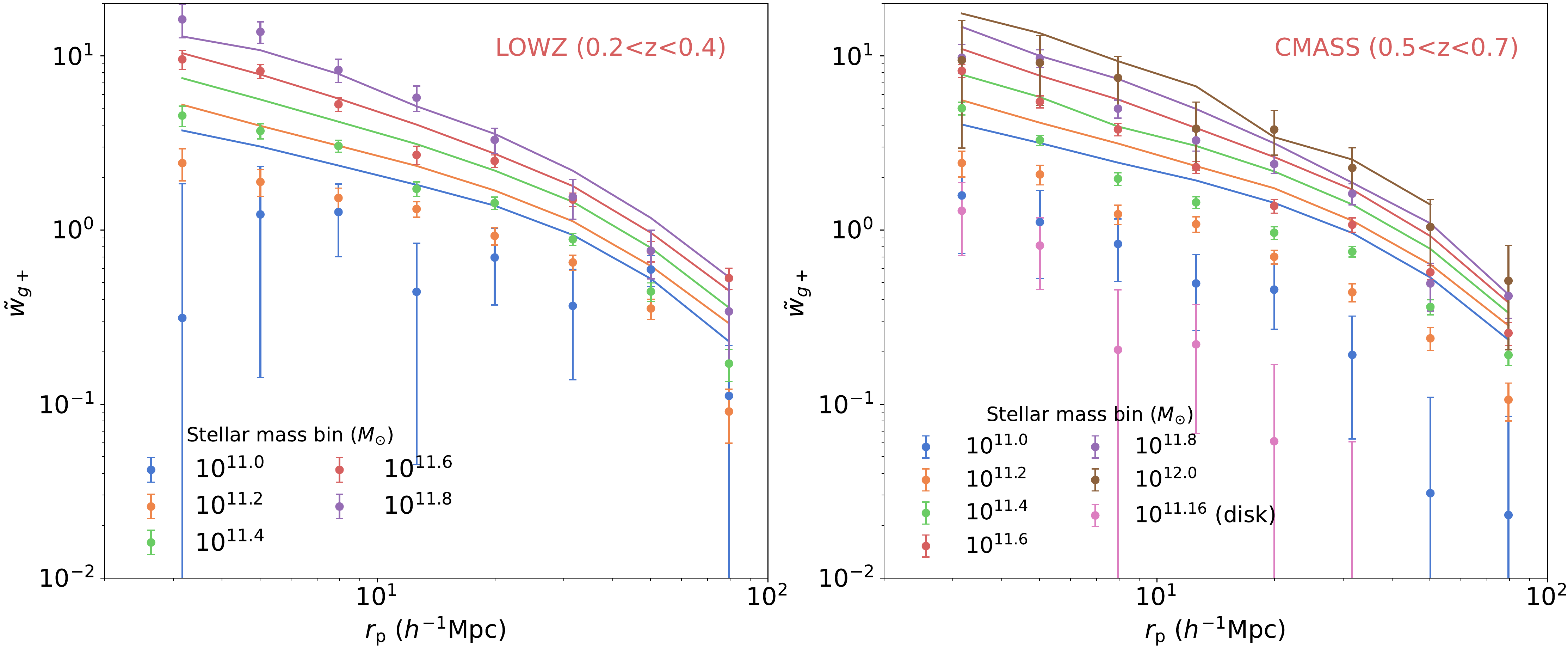}
    \caption{The GI correlations of central elliptical galaxies and their host DM halos in different stellar mass bins for the LOWZ (left) and CMASS (right) samples. Dots with error bars show the measurements for central elliptical galaxies and lines show the GI correlations for their host DM halos from simulations. The GI correlation of all the central disk galaxies from the CMASS sample is also shown for comparison.}
    \label{fig:fig0}
\end{figure*}

In this work, we focus on the galaxy-ellipticity (GI) correlation $\tilde{\xi}_{g+}$ since  $\tilde{\xi}_{g\times}$ should be 0 without parity-violating systematics \citep{2009ApJ...694L..83O}. Here we use $\tilde{\xi}_{g+}$ instead of $\xi_{g+}$ to represent the $q=0$ GI correlations. The GI correlation is defined as 
\begin{equation}
    \tilde{\xi}_{g+}(\bm{r})=\langle[1+\delta_{g_1}(\bm{x}_1)][1+\delta_{g_2}(\bm{x}_2)]e_+(\bm{x}_2)\rangle\,\,,
\end{equation}
where $\bm{r}=\bm{x}_1-\bm{x}_2$. The GI correlation can be estimated using the generalized Landy–Szalay estimator
 \citep{1993ApJ...412...64L,2006MNRAS.367..611M} with two random samples $R_s$ and $R$ corresponding to the tracers of ellipticity and density fields respectively,
\begin{equation}
    \tilde{\xi}_{g+}(r_{{\rm{p}}},\Pi) = \frac{S_{+}(D-R)}{R_{s}R}\,\,,
\end{equation}
where $R_sR$ is the normalized counts of random–random pairs in
a particular bin in the space of $(r_{{\rm{p}}},\Pi)$. $S_+D$ is the sum of the $+$ component of ellipticity in all pairs:
\begin{equation}
    S_+D=\sum_{i,j|r_{{\rm{p}}},\Pi}\frac{e_+(j|i)}{2\mathcal{R}}\,\,,
\end{equation}\label{eq:3}
where the ellipticity of the $j$th galaxy in the ellipticity tracers is defined relative to the direction to the $i$th galaxy in the density tracers, and $\mathcal{R}=1-\langle e_+^2 \rangle$ is the shape responsivity \citep{2002AJ....123..583B}. $\mathcal{R}$ equals to $0.5$ under our assumption of $q=0$. $S_+R$ is calculated in a similar way using the random catalog. Finally, the projected GI correlation function can be obtained,
\begin{equation}
    \tilde{w}_{g+}=\int_{-\Pi_{{\rm{max}}}}^{\Pi_{{\rm{max}}}}\tilde{\xi}_{g+}(r_{{\rm{p}}},\Pi)d\Pi\,\,.
\end{equation}
We adopt $\Pi_{{\rm{max}}}=80\ h^{-1}{\rm{Mpc}}$. We find that varying $\Pi_{{\rm{max}}}$ from $60\ h^{-1}{\rm{Mpc}}$ to $100\ h^{-1}{\rm{Mpc}}$ does not significantly change the results. 

We use central ellipticals as the tracers of the ellipticity field and split them into several stellar mass bins with an equal logarithmic interval of 0.2. In order to get better measurements, we use {\textit{all}} galaxies in LOWZ or CMASS samples as the density field tracers,  for all the stellar mass bins of ellipticity tracers. 

The GI measurements for central elliptical galaxies are shown in Figure \ref{fig:fig0}. Error covariance for GI correlation is estimated using jackknife resampling with 200 sub-samples. We can get relative good measurements in the stellar mass ranges of $[10^{10.9},10^{11.9}]M_{\odot}$ for LOWZ and $[10^{10.9},10^{12.1}]M_{\odot}$ for CMASS. We also measure the  GI correlation for all the central disk galaxies in the CMASS sample that have a mean stellar mass of $10^{11.16}M_{\odot}$. The result is shown in the figure (purple dots). The correlation of disk galaxies is much lower than that of the ellipticals, indicating that disk galaxies are much more misaligned with their host DM halos. 

\section{Modeling the misalignment angles}\label{sec:3}
As in \citet{2023ApJ...944..200X}, we use the {\texttt{CosmicGrowth}} simulations \citep{2019SCPMA..6219511J} to model the GI correlations and constrain the misalignment angles. We use the $\Lambda$CDM simulation with $3072^3$ dark matter particles in a cubic box of side $1200h^{-1}{\rm{Mpc}}$ and with cosmological parameters $\Omega_m = 0.268$, $\Omega_{\Lambda} = 0.732$ and $\sigma_8 = 0.831$. DM halos are found using the friends-of-friends (FOF) algorithm with a linking length of $b=0.2$ and are then processed with HBT+ \citep{2012MNRAS.427.2437H,2018MNRAS.474..604H} to find the subhalos and trace their evolution histories. 

We utilize the accurate ($\sim1\%$) stellar-halo mass relation (SHMR) measurements from \citet{2023ApJ...944..200X}, obtained using the Photometric Objects Around Cosmic Webs (PAC) method \citep{2022ApJ...925...31X}, to populate halos with galaxies. PAC measures the excess surface density $\bar{n}_2w_{{\rm{p}}}(r_{\rm{p}})$ of photometric objects with specific physical properties (such as stellar mass) around spectroscopic objects, without the need for photometric redshifts. Using the deep DESI Legacy Imaging Surveys, \citet{2023ApJ...944..200X} obtained 42 and 33 $\bar{n}_2w_{{\rm{p}}}(r_{\rm{p}})$ measurements for LOWZ and CMASS samples, respectively, with different spectroscopic and photometric stellar mass bins down to $10^{9.2}M_{\odot}$ and $10^{9.8}M_{\odot}$. These measurements are modeled using the subhalo abundance matching method to constrain the SHMR to within percent levels at both redshifts and all $\bar{n}_2w_{{\rm{p}}}(r_{\rm{p}})$ measurements are well fitted. The resulting galaxy stellar mass functions (GSMFs), derived from the model, are in good agreement with both model-independent measurements from PAC and measurements from photometric redshifts \citep{2022ApJ...939..104X}. The consistency among these three independent measurements of the GSMF supports the SHMR analysis results. 

We use the SHMRs of double power law forms at $0.2<z_s<0.4$ and $0.5<z_s<0.7$ from Table 2 of \citet{2023ApJ...944..200X} to generate mock density and ellipticity tracers. Halo and subhalo mass are defined as the virial mass $M_{{\rm{vir}}}$ of the halo and subhalo at the time when they was last the central dominant object. We use the fitting formula in \citet{1998ApJ...495...80B} to find $M_{{\rm{vir}}}$. All these definitions are consistent with that used in \citet{2023ApJ...944..200X} For the ellipticity tracers, since they are split to small stellar mass bins, we take all halos within the corresponding stellar mass bins. However, for the density tracers, which span a large range of stellar masses, we need to account for completeness at different stellar masses. To do this, we use the stellar mass completeness function from \citet{2023ApJ...944..200X}, which was derived by comparing the GSMF to the number densities of LOWZ and CMASS samples. After assigning centrals and satellites to halos and subhalos according to the SHMRs, We randomly reduce the number of density tracers in each stellar mass bin according to its completeness.  

\begin{figure}
    \plotone{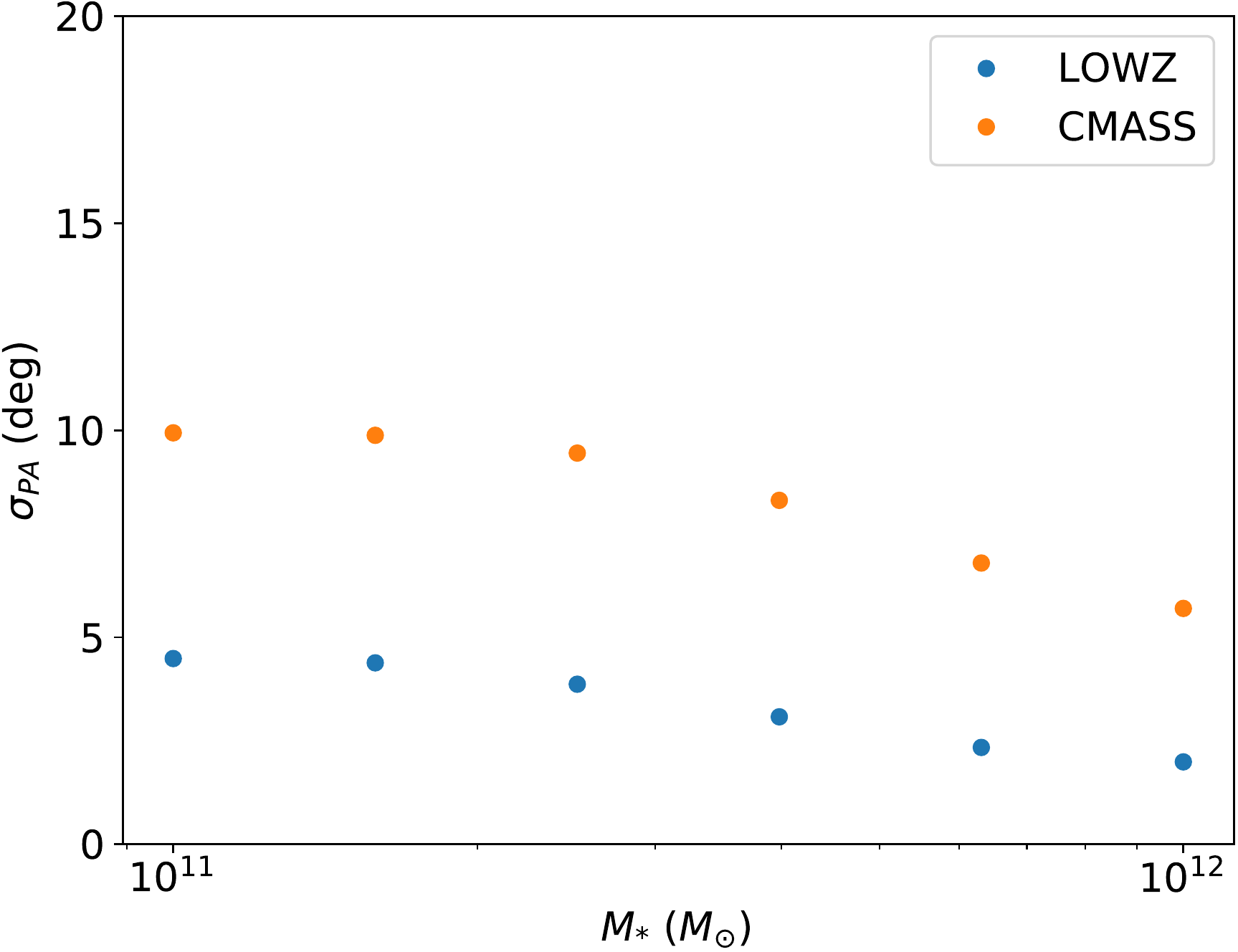}
    \caption{Mean errors of the position angles of galaxies in LOWZ and CMASS as a function of stellar mass.}
    \label{fig:PA}
\end{figure}

\begin{figure*}
    \plotone{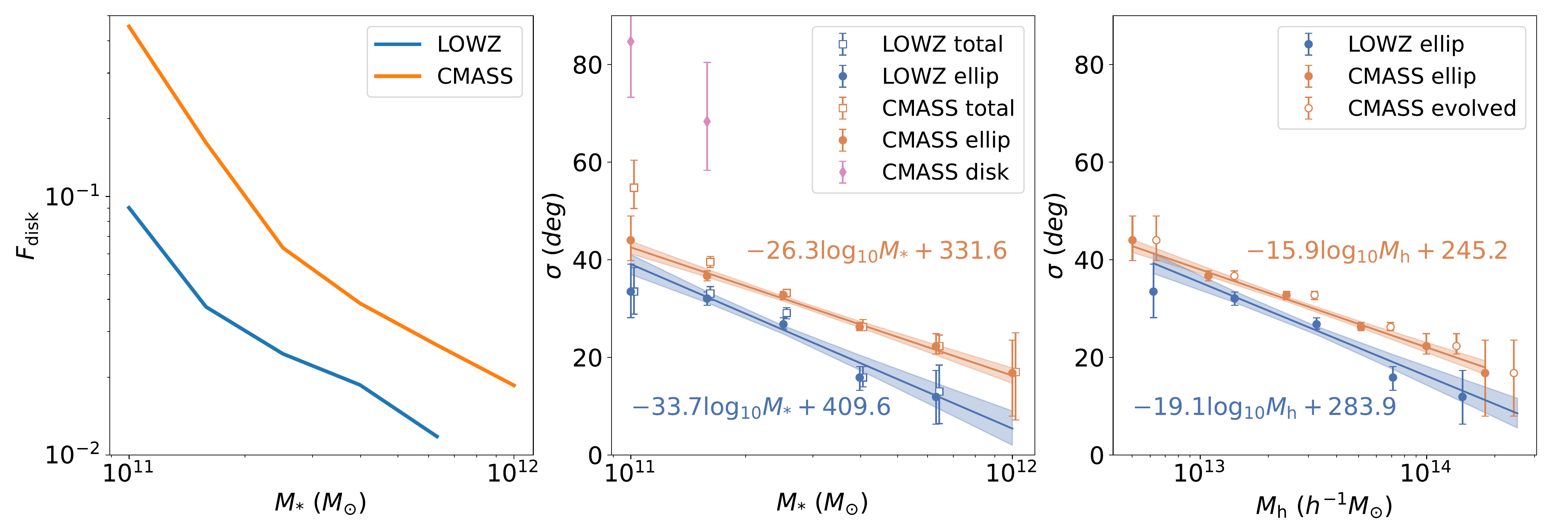}
    \caption{Left: the fraction of central disk galaxies as a function of stellar mass for the LOWZ and CMASS samples. Middle: the misalignment angle $\sigma_{\theta}$ as a function of stellar mass for central elliptical galaxies, central disk galaxies and all the central galaxies in both the LOWZ and CMASS samples. The best-fit relations for the central elliptical galaxies are shown with solid lines. Right: the misalignment angle $\sigma_{\theta}$ as a function of host halo mass for central elliptical galaxies. The best-fit relations are shown with solid lines. Open circles label the halo masses of the CMASS galaxies that are expected to grow to the LOWZ redshift, while the misalignment angles are assumed unchanged.}
    \label{fig:fig2}
\end{figure*}

\begin{figure*}
    \plotone{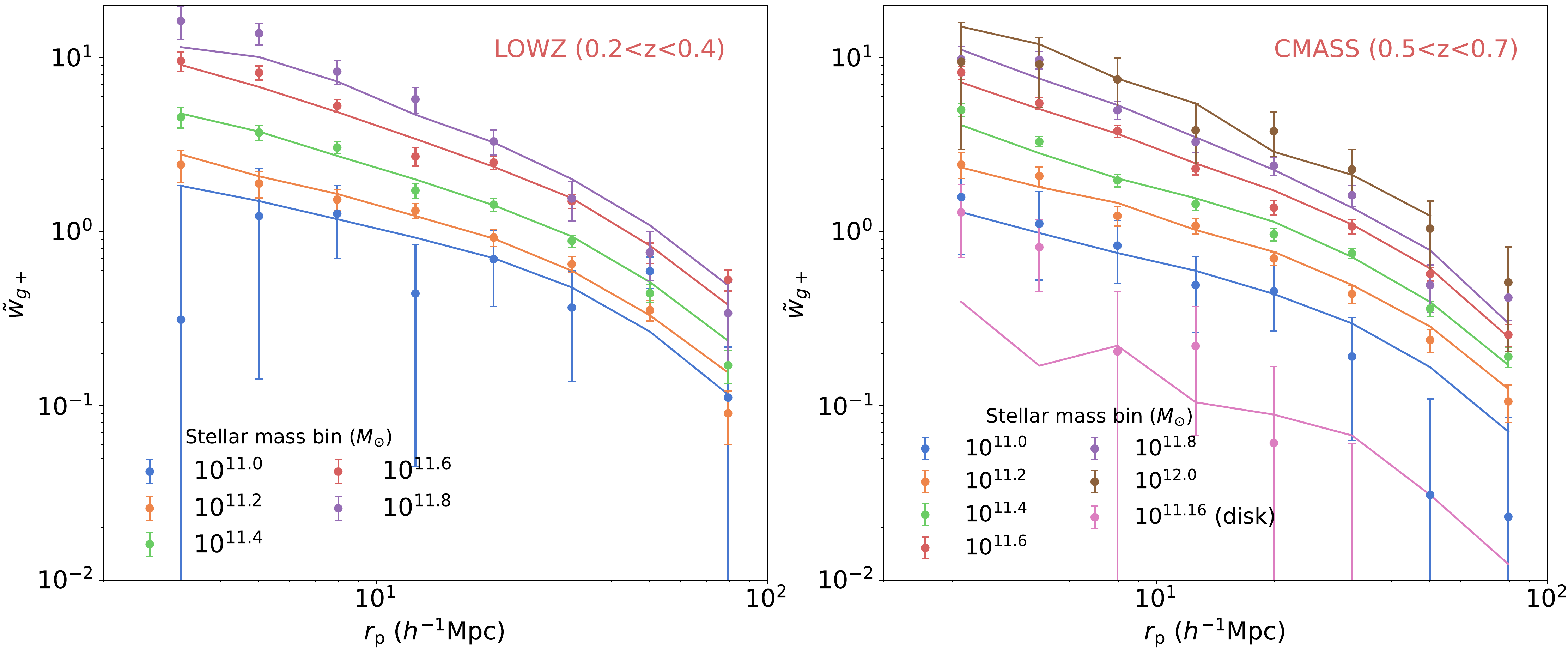}
    \caption{Similar to Figure \ref{fig:fig0}, the lines in this figure represent model predictions with the best-fit misalignment angles.}
    \label{fig:fig1}
\end{figure*}

The halo shapes of the ellipticity tracers are calculated using the iteration method \citep{1995MNRAS.276..417J}. We first use FOF algorithm with a linking length of 0.1 to select the central regions of the halos and then process them with the iteration method. Finally, we calculate their reduced 2D moments of inertia tensors $I_{ij}$ \citep{2005ApJ...627L..17B},
\begin{equation}
    I_{ij}=\sum_k\frac{x_{k,i}x_{k,j}}{x_k^2},\ {\rm{with}}\ i,j=\{1,2\}\,\,,
\end{equation}
where $x_k$ is the distance of the $k$th particle to the halo center. The major axes of the halos can be obtained by finding the eigenvectors and eigenvalues of $I_{ij}$. We also compare the results to those obtained using the non-reduced inertia tensors. We observe average deviations of $5^{\circ}$ to $10^{\circ}$ in the PA of halos between the two definitions in different bins. The GI correlations $\tilde{w}_{g+}$ are expected to be higher for the non-reduced version since it assigns greater weight to the outer region and should have a stronger alignment with the large-scale structure. However, the non-reduced version is more susceptible to substructures, leading to increased stochastic effects in PA, which in turn suppresses $\tilde{w}_{g+}$. These two effects nearly cancel each other out, and the amplitudes of $\tilde{w}_{g+}$ obtained from the two definitions are almost identical. As a result, both definitions exhibit nearly negligible differences in $\sigma_{\theta}$ ($1^{\circ}$ to $2^{\circ}$) when fitting the $\tilde{w}_{g+}$ values.

The GI correlations $\tilde{w}_{g+}$ of the host DM halos for each stellar mass bin are shown as lines in Figure \ref{fig:fig0}. It is observed that halos hosting larger central galaxies, and consequently possessing higher halo masses, exhibit higher $\tilde{w}_{g+}$ values, which is consistent with previous works \citep{2002MNRAS.335L..89J,2017ApJ...848...22X}.

Following \citet{2009ApJ...694L..83O}, we assume that the misalignment angle $\theta$ between the major axes of central galaxies and their host halos follows a Gaussian function with a zero mean and a width $\sigma_{\theta}$. We also consider the uncertainties of the position angles (PA) $\sigma_{{\rm{PA}}}$ in the measurements for each stellar mass bin as shown in Figure \ref{fig:PA}.
The errors in PA are determined from the model fitting performed by {\texttt{Tractor}}. As we will show below, $\sigma_{{\rm{PA}}}$ is much smaller than $\sigma_{\theta}$, and has a negligible effect on our results. With $\sigma_{{\rm{PA}}}$ and $\sigma_{\theta}$, for each stellar mass bin, we can get the orientations of the major axes of the mock ellipticity tracers and calculate the model predictions of the GI correlations. We calculate $\tilde{w}_{g+}$ for halos in each bins using a separation of $\Delta\sigma_{\theta}=0.1^{\circ}$ in the range of $0^{\circ}<\sigma_{\theta}<90^{\circ}$. For each value of $\sigma_{\theta}$, we perform the calculation of GI correlations 10 times using different random seeds for the misalignment angle distributions. We then obtain the final model prediction by averaging over the 10 GI correlations. We define the $\chi^2$ as
\begin{equation}
    \chi^2 = \sum_{a,b}\big[\tilde{w}_{g+,a}^{\mathrm{mod}}-\tilde{w}_{g+,a}^{\mathrm{obs}}\big]\mathbf{C}^{-1}_{ab}\big[\tilde{w}_{g+,b}^{\mathrm{mod}}-\tilde{w}_{g+,b}^{\mathrm{obs}}\big]\,\,,
\end{equation}
where $\mathbf{C}^{-1}$ is the inverse of the covariance matrix $\mathbf{C}$ and $a,b$ indicate the data points at different radial bins. We use the Markov chain Monte Carlo (MCMC) sampler {\texttt{emcee}} \citep{2013PASP..125..306F} to perform maximum likelihood analyses of \{$\sigma_{\theta}$\}.

The misalignment angles of central elliptical galaxies are shown in Figure \ref{fig:fig2} and the best-fit GI correlations are shown with solid lines in Figure \ref{fig:fig1}. The fits are overall good for all stellar mass bins. The misalignment angle $\sigma_{\theta}$ decreases linearly with $\log_{10}{M_{*}}$ for both LOWZ and CMASS samples as shown in the middle panel of Figure \ref{fig:fig2}, and central ellipticals at lower redshift are more aligned with their host halos at a fixed stellar mass,
\begin{equation}
\sigma_{\theta}=
\begin{cases}
-33.7^{+5.2}_{-5.3}\log_{10}M_{\ast}+409.6^{+60.0}_{-59.5}&\text{({\rm{LOWZ}})}\\
-26.3^{+2.7}_{-2.7}\log_{10}M_{\ast}+331.6^{+30.6}_{-31.0}&\text{({\rm{CMASS}})}\,\,.
\end{cases}
\end{equation}
We also show the dependence of $\sigma_{\theta}$ on the host halo mass in the right panel of Figure \ref{fig:fig2}, and we find
\begin{equation}
\sigma_{\theta}=
\begin{cases}
-19.1^{+3.0}_{-3.0}\log_{10}M_{{\rm{h}}}+283.9^{+40.3}_{-40.1}&\text{({\rm{LOWZ}})}\\
-15.9^{+2.7}_{-2.7}\log_{10}M_{{\rm{h}}}+245.2^{+22.5}_{-21.6}&\text{({\rm{CMASS}})}\,\,.
\end{cases}
\end{equation}
Similarly, we observe that at fixed halo mass, central ellipticals in LOWZ at lower redshifts are more aligned with their host halos. To check if this is caused by the halo mass growth, we also plot $\sigma_{\theta}$ (open circles) for the CMASS galaxies with the halo mass evolved to the LOWZ redshift using the subhalo merger trees (assuming no change of the misalignment). As we see, this is not the case as the halo mass growth would lead galaxies being more misaligned with host halos at a fixed halo mass. Thus, to match the LOWZ results, CMASS galaxies should have evolved to be more aligned with their host halos. Comparing the fitting slopes of LOWZ and CMASS, we find that the difference are within $2\sigma$ for $M_{{\rm{h}}}$ or $M_{\ast}$. More data is needed to see if the slopes do not evolve. The dependence of the misalignment angle on halo mass and redshift that we find is consistent with previous results from hydrodynamic simulations \citep{2014MNRAS.441..470T,2015MNRAS.453..721V,2017MNRAS.472.1163C}. While there appears to be an evolution among central elliptical galaxies at different redshifts, we should be careful about the different selections of the LOWZ and CMASS samples, since CMASS sample is bluer than LOWZ according to their target selection \citep{2016MNRAS.455.1553R,2022arXiv221211319S}. As shown in \citet{2023ApJ...944..200X}, LOWZ and CMASS samples are quite complete ($>75\%$) at $M_*>10^{11.3}M_{\odot}$ and $M_*>10^{11.5}M_{\odot}$ respectively, which may make the comparison reasonable at least for this mass range. However, there is still some incompleteness at the high mass end, better data is needed to confirm that the observed redshift dependence is not induced by selection effects.

In the middle panel of Figure \ref{fig:fig2}, we also show $\sigma_{\theta}$ for all the central galaxies. As expected, $\sigma_{\theta}$ is larger compared to that of only central ellipticals, especially for the lowest 2 stellar mass bins in the CMASS sample that have a larger disk fraction $F_{\rm{disk}}$ (left panel of Figure \ref{fig:fig2}). We further estimate $\sigma_{\theta}$ for central disk galaxies for these 2 stellar mass bins by simultaneously modeling the GI correlations of the central galaxies and of elliptical only central ones, with 2 different $\sigma_{\theta}$ for ellipticals and disks respectively. The modeled GI correlations for all the central galaxies are obtained by combining the modeled elliptical and disk GI correlations according to $F_{\rm{disk}}$. The results are also shown in the middle panel of Figure \ref{fig:fig2} (purple dots), we find that central disks are highly misaligned with their host halos ($\sigma_{\theta}>70^{\circ}$). The best-fit model for central disk galaxies with $10^{11.2}M_{\odot}$ (purple line in Figure \ref{fig:fig1}) is also consistent with our measurements for all the central disk galaxies in CMASS that have a mean stellar mass of $10^{11.16}M_{\odot}$. 

\section{Conclusion and discussion}\label{sec:4}
In this paper, we measure the GI correlations for massive central galaxies with different stellar masses at two different redshifts, and model them in an N-body simulation to get their galaxy-halo misalignment angles. 

We find that central elliptical galaxies with larger mass are more aligned with their host
halos, and the misalignment angle $\sigma_{\theta}$ depends linearly on the stellar mass $\log_{10}M_{\ast}$ and on their host halo mass $\log_{10}M_{{\rm{h}}}$ as well. We also find that central ellipticals are more aligned with their host halos in the lower redshift sample, which may indicate a evolution of the galaxy-halo alignment, though the differences between LOWZ and CMASS should be considered when interpreting these results. Furthermore, we find that central disk galaxies are highly misaligned ($\sigma_\theta>70^{\circ}$) with their host halos, with the GI correlation being about one magnitude weaker than their elliptical counterpart. 

These results are consistent with those obtained in state-of-the-art cosmological hydrodynamic simulations Illustris TNG300-1 \citep{2018MNRAS.480.5113M, 2018MNRAS.475..624N, 2018MNRAS.477.1206N,2018MNRAS.475..676S,2018MNRAS.475..648P,2019ComAC...6....2N} in a companion paper \citep{2023arXiv230712334X}, which helps us to better understand the results. In TNG300-1, We find that both the 2D and 3D misalignment angles for all principle axes decrease with ex situ stellar mass fraction $F_{\rm{acc}}$, halo mass $M_{\rm{h}}$ and stellar mass $M_{*}$, while increasing with disk-to-total stellar mass fraction $F_{\rm{disk}}$ and redshift, in which we find $F_{\rm{acc}}$ is a key factor that determine the galaxy-halo alignment. Grouping galaxies by $F_{\rm{acc}}$ nearly eliminates the dependence of misalignment angles on $M_{\rm{h}}$, and also reduces the redshift dependence. These results may help provide an explanation for the observed enhanced alignment in elliptical galaxies. Elliptical galaxies are primarily formed through mergers, which result in larger values of $F_{\rm{acc}}$, leading to a stronger alignment.


Although our results are largely consistent with previous studies, our accurate quantification of $\sigma_{\theta}$ can be used as an important test for galaxy formation models, since the mass and redshift dependences could be very sensitive to how massive galaxies have been assembled.
Our results can also benefit the weak lensing and IA cosmology studies. Combining the SHMR and the galaxy-halo misalignment results, one can generate realistic galaxy mocks with shape information. The mocks can be used to model the IA effect in weak lensing measurements, and give a forecast for IA cosmology. 

With the next generation of larger and deeper spectroscopic and photometric surveys, such as Dark Energy Spectroscopic Instrument \citep{2016arXiv161100036D}, Prime Focus Spectrograph \citep{2016SPIE.9908E..1MT}, China Space Station Telescope \citep{2018cosp...42E3821Z}, Legacy Survey of Space and Time \citep{2019ApJ...873..111I} and Euclid \citep{2011arXiv1110.3193L}, we will explore the galaxy-halo misalignment to higher redshift and lower mass. 

\section*{Acknowledgments}
The work is supported by NSFC (12133006, 11890691, 11621303), grant No. CMS-CSST-2021-A03, and 111 project No. B20019. We gratefully acknowledge the support of the Key Laboratory for Particle Physics, Astrophysics and Cosmology, Ministry of Education. This work made use of the Gravity Supercomputer at the Department of Astronomy, Shanghai Jiao Tong University.

Funding for SDSS-III has been provided by the Alfred P. Sloan
Foundation, the Participating Institutions, the National Science
Foundation, and the US Department of Energy Office of Science. The SDSS-III web site is http://www.sdss3.org/. 

The Legacy Surveys consist of three individual and complementary projects: the Dark Energy Camera Legacy Survey (DECaLS; Proposal ID \#2014B-0404; PIs: David Schlegel and Arjun Dey), the Beijing-Arizona Sky Survey (BASS; NOAO Prop. ID \#2015A-0801; PIs: Zhou Xu and Xiaohui Fan), and the Mayall z-band Legacy Survey (MzLS; Prop. ID \#2016A-0453; PI: Arjun Dey).


\bibliography{sample63}{}
\bibliographystyle{aasjournal}




\end{document}